\documentclass [twocolumn, aps, prb, superscriptaddress, showpacs] {revtex4}
\usepackage{graphicx,latexsym}
\begin{document}
\draft
\title {Coulomb Drag of Massless Fermions in Graphene}
\author {Seyoung Kim}
\affiliation {Microelectronics Research Center, The University of
Texas at Austin, Austin, TX 78758}
\author {Insun Jo}
\affiliation {Department of Physics, The University of Texas at
Austin, Austin, TX 78712}
\author {Junghyo Nah}
\affiliation {Microelectronics Research Center, The University of
Texas at Austin, Austin, TX 78758}
\author {Z. Yao}
\affiliation {Department of Physics, The University of Texas at
Austin, Austin, TX 78712}
\author {S. K. Banerjee}
\affiliation {Microelectronics Research Center, The University of
Texas at Austin, Austin, TX 78758}
\author {E. Tutuc}
\email {etutuc@mail.utexas.edu}
\affiliation {Microelectronics Research Center, The University of
Texas at Austin, Austin, TX 78758}
\date{\today}
\begin{abstract}
Using a novel structure, consisting of two, independently contacted graphene
single layers separated by an ultra-thin dielectric, we experimentally measure
the Coulomb drag of massless fermions in graphene.  At temperatures higher than
50 K, the Coulomb drag follows a temperature and carrier density dependence consistent with
the Fermi liquid regime.  As the temperature is reduced, the Coulomb drag exhibits
giant fluctuations with an increasing amplitude, thanks to the interplay between
coherent transport in the  graphene layer and interaction between the two layers.
\end{abstract}
\pacs{73.43.-f, 71.35.-y, 73.22.Gk} \maketitle

Bilayer systems formed by two layers of carriers in close proximity are a fascinating
testground for electron physics. In particular, the prospect of electron-hole pair (indirect exciton)
formation, and dipolar superfluidity \cite{lozovik} has fueled the research of electron-hole
bilayers in GaAs/AlGaAs heterostructures \cite{pohlt,croxall}. Graphene \cite{novoselov04,novoselov05}
is a particularly interesting material to explore interacting bilayers. The symmetric conduction and valence bands,
and the large Fermi energy favor correlated electron states at elevated temperatures \cite{min,zhang}.
The zero energy band-gap allows a seamless transition between electrons and holes in each layer,
and obviates the large inter-layer electric field required to simultaneously induce electrons and holes in
GaAs bilayers \cite{croxall}. Coulomb drag, a direct measurement of inter-layer electron-electron
scattering \cite{solomon} can provide insight into the ground state of two- \cite{gramila} and one- \cite{yamamoto}
electron systems, as well as correlated bilayer states \cite{vignale,tutuc}. Here we demonstrate a novel,
independently contacted graphene bilayer, and investigate the Coulomb drag in this system.

Two main ingredients render the realization of independently contacted graphene bilayers challenging.
First, an ultra-thin yet highly insulating dielectric is required to separate the two layers.
Second, a method to position another graphene layer on a pre-existing device with minimum or no degradation is needed to create the second
layer of the structure investigated here. The fabrication of our independently contacted graphene bilayers
is described in Fig. 1. First, the bottom graphene layer is mechanically exfoliated onto a 280 nm thick SiO$_2$ dielectric,
thermally grown on a highly doped Si substrate. E-beam lithography, metal lift-off, and etching are used to define a Hall bar on the bottom layer [Fig. 1(a)].
A 7 nm thick Al$_2$O$_3$ is then deposited on the bottom layer using a 2 nm oxidized Al interfacial layer,
followed by 5 nm of Al$_2$O$_3$ atomic layer deposition \cite{kim}. The second, top graphene layer is also
mechanically exfoliated on a similar SiO$_2$/Si substrate. A poly methyl metacrylate (PMMA) film is applied on the top layer and cured.
Using an NaOH etch \cite{reina}, the PMMA film along with the graphene layer,
and the alignment marks are detached from the host substrate, forming a free standing membrane. The membrane is
placed face down on the substrate containing the bottom graphene layer [Fig. 1(b)], and aligned with it.
A Hall bar is subsequently defined on the top layer [Fig. 1(c)]. Ten back-gated, independently
contacted graphene bilayers have been fabricated and investigated in this study, all with similar results.
We focus here on data collected from three samples, labeled 1, 2, and 3, with mobilities between
4,200-12,000 cm$^2$/Vs for the bottom layer, 4,500-22,000 cm$^2$/Vs for the top layer, and with
inter-layer resistances of 1-20 G$\Omega$. These structures are markedly different from
graphene bilayers exfoliated from natural graphite, consisting of two graphene monolayers in Bernal stacking \cite{mccann}.

\begin{figure*}
\centering
\includegraphics[scale=0.4]{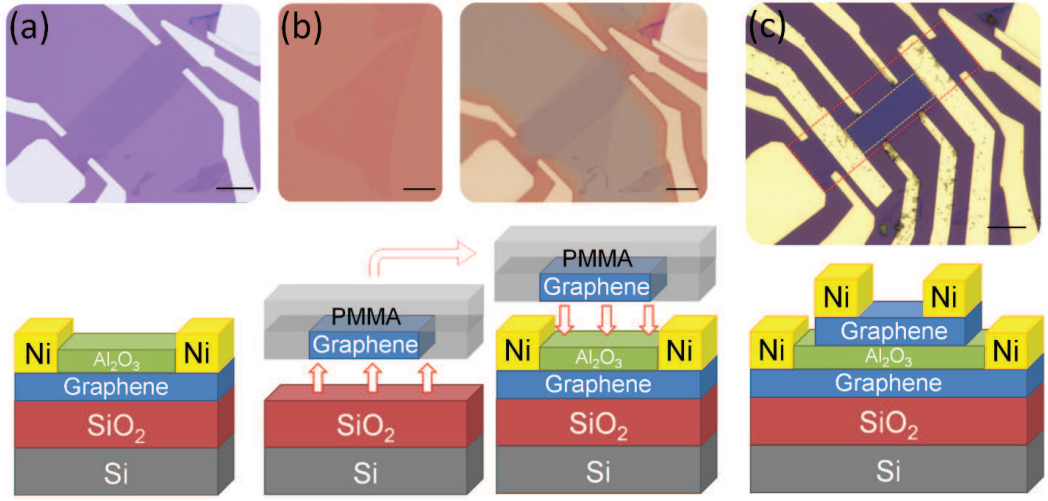}
\caption {\small{(color online) Optical micrographs (top) and schematic representation (bottom)
of the fabrication process of an independently contacted graphene bilayer.  (a) Hall bar
device fabrication on the bottom graphene layer, followed by the Al$_2$O$_3$ inter-layer dielectric deposition.
(b) Top graphene layer isolation on a separate substrate, followed by transfer onto the bottom layer. (c)
Top layer Hall bar realization by etching, lithography, metal deposition, and lift-off.
The yellow (inner) and red (outer) dashed contours in the optical micrograph represent the top and bottom layers, respectively.
The scale bars in all panels are 10 $\mu$m.}}
\end{figure*}

We now turn to the individual layer characterization. The layer resistivities ($\rho$) and Hall densities measured for sample 1 at a temperature T=4.2 K,
as a function of back-gate bias ($V_{BG}$) are shown in Fig. 2(a) and 2(b), respectively.
The potential of the both layers is held at zero (ground) for all measurements presented in this study.
The bottom layer dependence on the applied $V_{BG}$ shows ambipolar conduction and a finite resistance
at the charge neutrality (Dirac) point, consistent with the expected response of gated monolayer graphene
\cite{adam}. More interestingly, the top layer resistivity also changes as a result of the applied $V_{BG}$.
This observation indicates an incomplete screening of the gate-induced electric field by the bottom layer \cite{schmidt}, which is most pronounced in the
vicinity of the charge neutrality point, consequence of the reduced density of states in graphene.
As we show below, we can quantitatively explain the layer resistivities and densities dependence on $V_{BG}$.

\begin{figure*}
\centering
\includegraphics[scale=0.38]{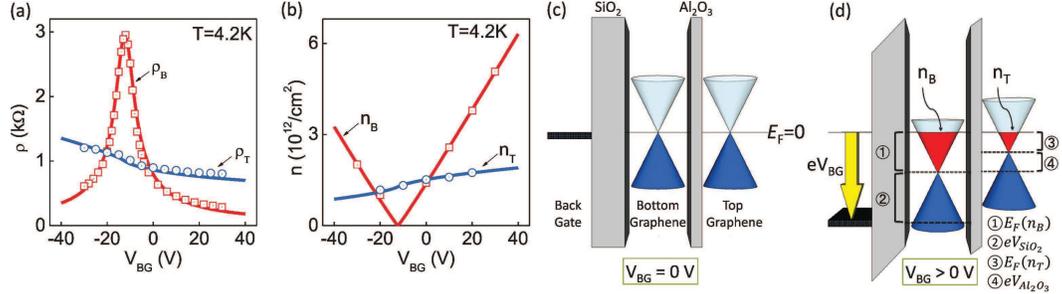}
\caption {\small{(color online) (a) Layer resistivities, and (b) densities vs. $V_{BG}$ measured at $T$=4.2 K in sample 1.
Depending on $V_{BG}$, both electrons or holes can be induced in the bottom layer;
the top layer contains electrons in the available $V_{BG}$ window, owing to unintentional doping.
The symbols in panels (a) and (b) represent experimental data, while the lines represent the calculated values
according to Eqs. (1) and (2). (c) Band diagram across the graphene bilayer heterostructure at $V_{BG}$=0 V, and (d) at a positive $V_{BG}$.
Both layers are assumed to be at the charge neutrality point, and aligned with the back-gate Fermi level at $V_{BG}=0$.
The layers are held at the ground potential, and their thicknesses exaggerated to show the Dirac cones.
The applied $V_{BG}$ induces voltage drops $V_{SiO_{2}}$, and $V_{Al_{2}O_{3}}$ across the bottom, and inter-layer dielectrics
respectively.}}
\end{figure*}

Figure 2(c) shows the band diagram of the graphene bilayer at $V_{BG}$=0 V; for simplicity the gate Fermi energy
and the charge neutrality point in the two layers are assumed to be at the same energy. Once a finite $V_{BG}$
is applied, finite charge densities are induced in both top ($n_T$) and bottom ($n_B$) layers [Fig. 2(d)].
The difference between the gate and bottom layer Fermi levels is distributed partly across the SiO$_2$ dielectric,
and partly on the Fermi energy of the bottom graphene layer: $eV_{BG}=e^2(n_B+n_T)/C_{SiO_2}+E_F(n_B)$ (1);
$E_F(n)=\hbar v_F \sqrt{\pi n}$ is the graphene Fermi energy measured with respect to the charge neutrality
point at a carrier density $n$, $e$ is the electron charge, $v_F$=1.1$\times10^6$ m$/$s is the graphene Fermi velocity,
and $C_{SiO_2}$ denotes the SiO$_2$ dielectric capacitance per unit area. Similarly, the Fermi energy difference between
the two layers is responsible for the potential drop across the Al$_2$O$_3$ inter-layer dielectric: $E_F(n_B)=e^2n_T/C_{Al_2O_3}+E_F(n_T)$ (2);
$C_{Al_2O_3}$ is the Al$_2$O$_3$ dielectric capacitance per unit area. The finite Fermi energy of the bottom layer, $E_F(n_B)$ in Eq. (2),
plays the same role with respect to the top layer, as the applied $V_{BG}$ in Eq. (1) for the bottom layer.
Equations (1) and (2) allow us to determine $n_B$ and $n_T$ as a function of $V_{BG}$. This model can be adjusted
to include finite layer densities at $V_{BG}$=0 V. The layer resistivity dependence on $V_{BG}$ can be understood using a Drude model $\rho_{T,B}=(n^*_{T,B}e\mu_{T,B})^{-1}$, where $\mu_T$, and $\mu_B$ are the top and bottom layer mobilities,
and the layer densities $n_{T,B}=\sqrt{n_{T,B}^2+n_{0T,0B}^2}$ are adjusted to allow for finite carrier densities ($n_{0T}$, $n_{0B}$) at the charge neutrality point.
The data of Figs. 2(a) and 2(b) show a good agreement between the measured layer resistivities and densities (symbols),
and the calculations (solid lines). The layer mobilities, determined from Hall measurements, are  $\mu_B$=5,400 cm$^2$/Vs,
and $\mu_T$=4,500 cm$^2$/Vs at $T$=4.2 K.

\begin{figure}
\centering
\includegraphics[scale=0.38]{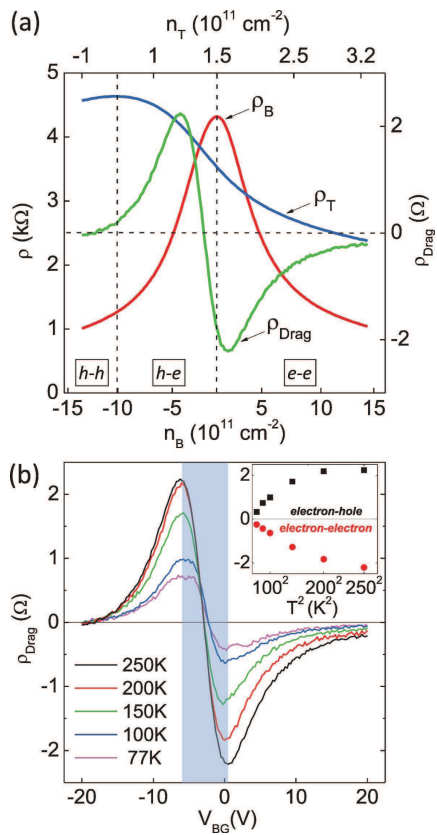}
\caption {\small{(color online) Coulomb drag in graphene. (a) Layer resistivities ($\rho_{T,B}$) and $\rho_{Drag}$ vs. layer densities ($n_{T,B}$) for sample 2,
measured by sweeping $V_{BG}$ at $T$=250 K. The bilayer probes three different regimes: hole-hole, electron-hole, and electron-electron.
(b) $\rho_{Drag}$ vs. $V_{BG}$ at different $T$ values, from 250 K to 77 K (solid lines). Inset: maximum $\rho_{Drag}$ vs. $T^2$ in the electron-hole and electron-electron regimes. The different $x$-axis, i.e. $n_B$ and $n_T$ of panel (a) and $V_{BG}$ of panel (b), apply to both panels.}}
\end{figure}

Key insight into the physics of the graphene bilayer system can be gained from Coulomb drag measurements \cite{solomon,gramila}.
A current ($I_{Drive}$) flown in one (drive) layer leads to a momentum transfer between the two layers, thanks to the
inter-layer electron-electron interaction. To counter this momentum transfer, a longitudinal voltage
($V_{Drag}$) builds up in the opposite (drag) layer. The polarity of $V_{Drag}$ depends on the carrier
type in the two layers, and is opposite (same) polarity as the voltage drop in the drive layer when
the both layers have the same (opposite) type of carriers. The drag resistivity is defined as
$\rho_{Drag}=(W/L)V_{Drag}/I_{Drive}$, where $L$ and $W$ are the length and width of the region where drag occurs.
$\rho_{Drag}$ vs. $V_{BG}$ measured at $T$=250 K in sample 2 is shown in Fig. 3(a),
along with the layer resistivities, $\rho_T$ and $\rho_B$. Unlike sample 1 data (Fig. 2),
the charge neutrality (Dirac) points of both layers can be captured in the experimentally accessible $V_{BG}$ window.
Consequently, depending on the $V_{BG}$ value, sample 2 can probe three different regimes: a {\it hole-hole} bilayer, for $V_{BG}$ $<$ -15 V,
an {\it electron-hole} bilayer for -15 V $<$ $V_{BG}$ $<$ -2 V, and an {\it electron-electron} bilayer for $V_{BG}$ $>$ -2 V.
The dependence of $\rho_B$ and $\rho_T$ on $V_{BG}$ of Fig. 3 is also in good agreement with the model presented in Fig. 2.
Consistent with the above argument, $\rho_{Drag}$ is positive in the electron-hole bilayer regime,
negative in the hole-hole or electron-electron regime, and changes sign when either the top or the bottom
layer are at the charge neutrality point. Standard consistency checks \cite{gramila} ensured
the measured drag is not affected by inter-layer leakage current.

\begin{figure}
\centering
\includegraphics[scale=0.17]{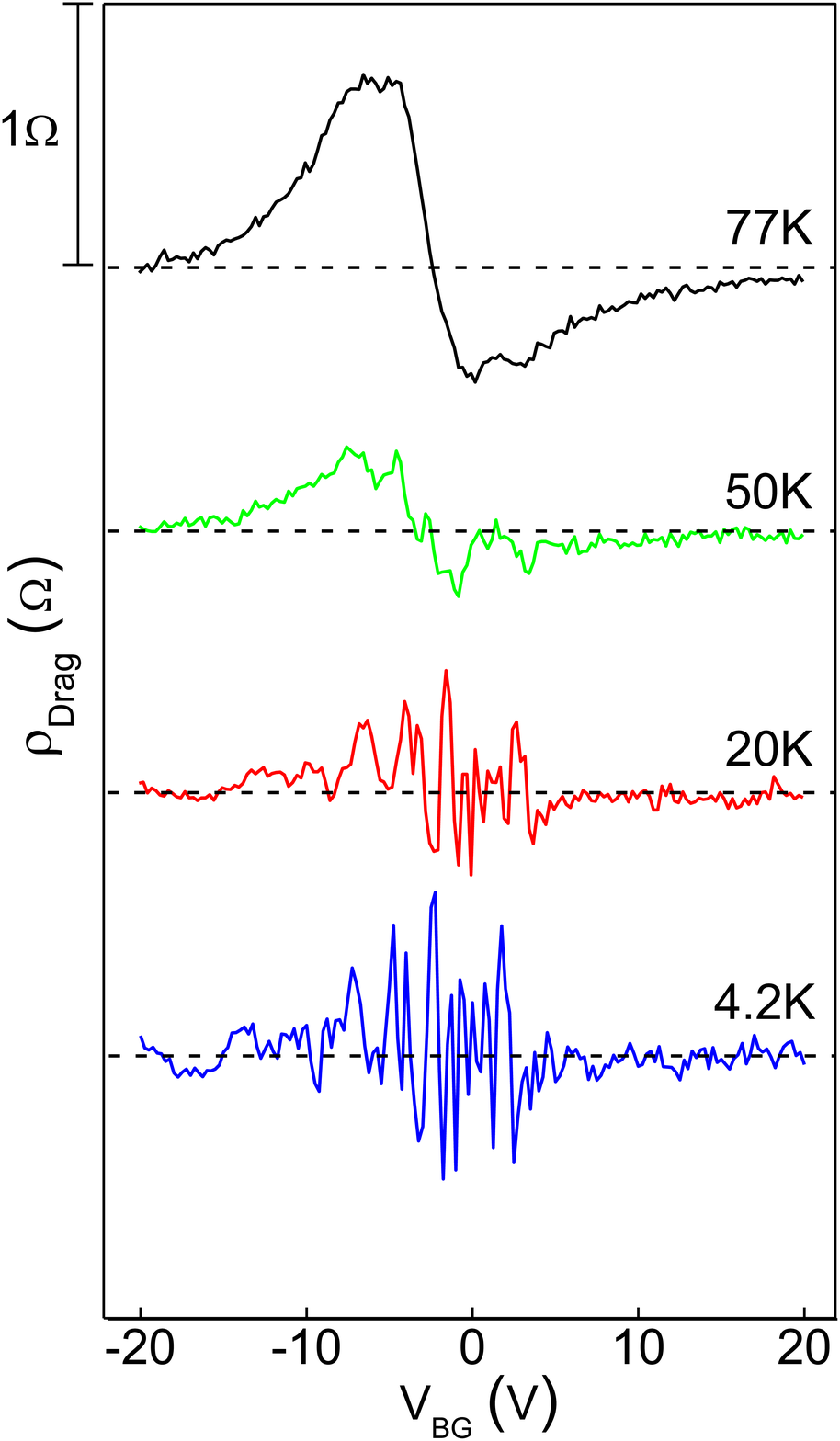}
\caption {\small{(color online) $\rho_{Drag}$ vs. $V_{BG}$ measured in sample 2 for $T \leq 77$ K.
As $T$ is reduced, $\rho_{Drag}$ exhibits mesoscopic fluctuations with increasing amplitude,
which fully obscure the average drag at the lowest $T$. The traces are shifted for clarity;
the horizontal dashed lines indicate 0 $\Omega$ for each trace.}}
\end{figure}

For two closely spaced two-dimensional systems, when the ground state of each layer is a Fermi liquid,
and the inter-layer interaction is a treated as a perturbation, the $\rho_{Drag}$ depends on layer density ($n$) as $\propto 1/n^{3/2}$,
on temperature as $\propto T^{2}$, and inter-layer distance ($d$) as $\propto 1/d^{4}$. \cite{gramila} Likewise, the Coulomb drag resistivity
in graphene, calculated in the Fermi liquid regime using Boltzmann transport formalism and the random phase approximation for the
dynamic screening is \cite{tse}: $\rho_{Drag}=-\frac{h}{e^6} \frac{\zeta(3)}{32} \frac{(k_{B}T)^2}{d^4} \frac{\epsilon ^2}{n_{B}^{3/2}n_{T}^{3/2}}$ (3);
$k_B$ is the Boltzmann constant, $\zeta (3)\cong 1.2$, $\epsilon$ and is the dielectric permittivity.
A separate effect which has been theoretically advanced as the representative Coulomb drag mechanism in graphene is trigonal warping \cite{narozhny}.
Figure 3(b) data shows $\rho_{Drag}$ vs. $V_{BG}$ measured for sample 2 for $T$ values
between 77 K and 250 K. Away from the bottom layer charge neutrality point, the $\rho_{Drag}$ magnitude decreases with increasing $n_B$ and $n_T$.
A power law, $\rho_{Drag}\propto 1/(n_B^{\alpha}n_T^{\alpha})$ fitting to Fig. 3 data for $V_{BG} > 0$ yields an exponent $\alpha$ = 1.25 $\pm 0.25$,
which depends little on temperature. We note that the magnitude of $\rho_{Drag}$ is a factor of $\sim 10^2$ lower than the values expected according to Eq. (3).
While further theoretical work is needed to explain this discrepancy, a possible explanation is that Eq. (3) is valid for high densities and/or large inter-layer spacing such that $k_F \cdot d \gg1$ ($k_F$ denotes the Fermi wave-vector) \cite{tse}; for Fig. 3 data $k_F\cdot d\leq3$ at all layer densities. The $\rho_{Drag} \propto (k_{B}T)^2$ dependence, which stems from the allowed phase space where electron-electron scattering occurs, is followed closely for temperatures between 70 K and 200 K (Fig. 3(b) inset), and softens for $T>200$ K. Figure 3 data shows a smooth crossover for $\rho_{Drag}$ through 0 $\Omega$, from the electron-hole to the electron-electron regime [blue (shaded) corridor of Fig. 3(b)]. The crossover can be explained by the co-existence of electron and hole puddles near the charge neutrality point of the bottom layer, which generate drag electric fields of opposite sign, and cancel the $\rho_{Drag}$.

\begin{figure}
\centering
\includegraphics[scale=0.17]{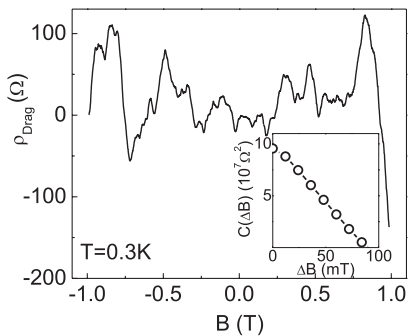}
\caption {\small{$\rho_{Drag}$ vs. $B$ measured at $T$=0.3 K in sample 3,
showing mesoscopic fluctuations similar to Fig. 4 data. Inset: $\rho_{Drag}$ vs. $B$ data
autocorrelation.}}
\end{figure}

A remarkable transition in the drag resistance is observed for $T$ lower than 50 K (Fig. 4). As $T$ is reduced,
the $\rho_{Drag}$ data starts to develop fluctuations superposed on the average $\rho_{Drag}$ vs. $V_{BG}$
dependence of Fig. 3 and Eq.(3), valid for diffusive transport. The $\rho_{Drag}$ fluctuations, which are reproducible
in different measurements, grow in amplitude as $T$ is reduced, and fully obscure the average diffusive drag below 20 K.
This manifestation of mesoscopic physics at elevated temperatures is a consequence of the phase coherence length ($L_{\varphi}$)
increasing with reducing $T$, and represents the counterpart of universal conductance fluctuations \cite{lee}
in Coulomb drag \cite{narozhny2000,price}. Figure 4 data reveal that $\rho_{Drag}$ fluctuation amplitude reaches
a maximum near the charge neutrality point of the bottom layer ($V_{BG}=-1$ V), and increases, albeit slowly as $T$ is decreased.
Theoretical arguments \cite{price} indicate that the drag conductivity ($\sigma_{Drag}= \rho_{Drag}/{\rho_T}{\rho_B}$) fluctuation amplitude ($\delta\sigma_{Drag}$)
depends on relevant length scales and temperature as $\delta\sigma_{Drag}\propto T \cdot (L_{\varphi}^3l)/L$; $l$ is the electron mean free path.
For the temperature range examined in Fig. 4, $l$ can be considered constant, as the mobility is weakly dependent on $T$.
Assuming the electron-electron interaction is the main phase-breaking mechanism in graphene \cite{altshuler},
hence $L_{\varphi}=l\sqrt{E_F/2k_BT}$, the $T$ dependence of $\delta\sigma_{Drag}$ and $\delta\rho_{Drag}$ should follow a $\propto T^{-1/2}$
dependence, in good agreement with Fig. 4 data.

To probe the signature of weak localization in Coulomb drag, in Fig. 5 we show an example of $\rho_{Drag}$ vs. perpendicular magnetic field ($B$)
data, measured in sample 3 at $T$=0.3 K and $V_{BG}$=0 V; both layers contain electrons with layer densities $n_{T}=1.4\times10^{11}$ cm$^{-2}$, and $n_{B}=1.5\times10^{11}$ cm$^{-2}$. Similar to the $V_{BG}$ dependence of Fig. 4, the $\rho_{Drag}$ vs. $B$ data shows reproducible mesoscopic fluctuations.
The auto-correlation function ($C(\Delta B)$) of Fig. 5 data reveals a correlation field $B_c=47$ mT, which corresponds to a phase coherence length $L_{\varphi}=\sqrt{h/eB_c}=300$ nm. Similar $L_{\varphi}$ values have been extracted from ensemble average measurements using scanning gate microscopy \cite{berezovsky}.

In summary, we demonstrate independently contacted graphene bilayers, and probe the Coulomb drag in this system.
At elevated temperatures the drag resistance dependence on density and temperature are consistent with the Fermi
liquid theory. At reduced temperatures, the drag exhibits mesoscopic fluctuations which obscure the average drag,
a result of the interplay between electron-electron interaction and phase coherent transport.

We thank A. H. MacDonald, W. K. Tse, and B. Narozhny for discussions, and  NRI-SWAN for support.
Part of our work was performed at the National High Magnetic Field Laboratory, which is supported
by NSF (DMR--0654118), the State of Florida, and DOE.


\begin{thebibliography}{10}
\small

\bibitem{lozovik}
Y. E. Lozovik, V. I. Yudson, JETP Lett. {\bf 22}, 274 (1975).

\bibitem{pohlt}
M. Pohlt {\it et~al.}, Appl. Phys. Lett. {\bf 80}, 2105 (2002).

\bibitem{croxall}
A. F. Croxall {\it et~al.}, Phys. Rev. Lett. {\bf 101}, 246801 (2008);
J. A. Seamons, C. P. Morath, J. L. Reno, M. P. Lilly, Phys. Rev. Lett. {\bf 102}, 026804 (2009).

\bibitem{novoselov04}
K. S. Novoselov {\it et~al.}, Science {\bf 306}, 666 (2004).

\bibitem{novoselov05}
K. S. Novoselov {\it et~al.}, Nature {\bf 438}, 197 (2005); Y.
Zhang {\it et~al.}, Nature {\bf 438}, 201 (2005).

\bibitem{min}
H. Min, R. Bistritzer, J.-J. Su, A. H. MacDonald, Phys. Rev. B {\bf 78}, 121401 (2008).

\bibitem{zhang}
C.-H. Zhang, Y. N. Joglekar, Phys. Rev. B {\bf 77} 233405 (2008).

\bibitem{solomon}
P. M. Solomon, P. J. Price, D. J. Frank, D. C. La Tulipe, Phys. Rev. Lett. {\bf 63}, 2508 (1989).

\bibitem{gramila}
T. J. Gramila, J. P. Eisenstein, A. H. MacDonald, L. N. Pfeiffer, K. W. West, Phys. Rev. Lett. {\bf 66}, 1216 (1991).

\bibitem{yamamoto}
M. Yamamoto, M. Stopa, Y. Tokura, Y. Hirayama, S. Tarucha, Science {\bf 313},
204 (2006).

\bibitem{vignale}
G. Vignale, A. H. MacDonald, Phys. Rev. Lett. {\bf 76}, 2786 (1996).

\bibitem{tutuc}
E. Tutuc, R. Pillarisetty, M. Shayegan, Phys. Rev. B. {\bf 79}, 041303(R) (2009).

\bibitem{kim}
S. Kim {\it et~al.}, Appl. Phys. Lett. {\bf 94}, 062107 (2009).

\bibitem{reina}
A. Reina {\it et~al.}, J. Phys. Chem. C {\bf 112}, 17741 (2008).

\bibitem{mccann}
E. McCann, V. I. Fal'ko, \prl {\bf 96}, 086805 (2006).

\bibitem{adam}
S. Adam, E. H. Hwang, V. M. Galitski, S. Das Sarma, Proc. Natl. Acad. Sci. U.S.A. {\bf 104}, 18392 (2007).

\bibitem{schmidt}
H. Schmidt {\it et~al.}, Appl. Phys. Lett. {\bf 93}, 172108 (2008).

\bibitem{tse}
W.-K. Tse, B. Y.-K. Hu, S. Das Sarma, Phys. Rev. B {\bf 76}, 081401 (2007).

\bibitem{narozhny}
B. N. Narozhny, Phys. Rev. B {\bf 76}, 153409 (2007).

\bibitem{lee}
P. A. Lee, A. D. Stone, Phys. Rev. Lett. {\bf 55}, 1622 (1985).

\bibitem{narozhny2000}
B. N. Narozhny, I. L. Aleiner, Phys. Rev. Lett. {\bf 84}, 5383 (2000).

\bibitem{price}
A. S. Price, A. K. Savchenko, B. N. Narozhny, G. Allison, D. A. Ritchie, Science {\bf 316}, 99 (2007).

\bibitem{altshuler}
B. L. Altshuler, A. G. Aronov, Electron-electron interactions in disordered systems (North-Holland, Amsterdam, 1985).

\bibitem{berezovsky}
J. Berezovsky, R. M. Westervelt, Nanotechnology {\bf 21}, 274014 (2010).

\end{thebibliography}
\end{document}